# WANA: Symbolic Execution of Wasm Bytecode for Cross-Platform Smart Contract Vulnerability Detection[*][#]

Dong Wang, Bo Jiang[†]
School of Computer Science and Engineering
Beihang University
Beijing, China
{wangdong, jiangbo}@buaa.edu.cn

W.K. Chan
Department of Computer Science
City University of Hong Kong
Hong Kong
wkchan@cityu.edu.hk

## ABSTRACT

Many popular blockchain platforms are supporting smart contracts for building decentralized applications. However, the vulnerabilities within smart contracts have led to serious financial loss to their end users. For the EOSIO blockchain platform, effective vulnerability detectors are still limited. Furthermore, existing vulnerability detection tools can only support one blockchain platform. In this work, we present WANA, a cross-platform smart contract vulnerability detection tool based on the symbolic execution of WebAssembly bytecode. Furthermore, WANA proposes a set of test oracles to detect the vulnerabilities in EOSIO and Ethereum smart contracts based on WebAssembly bytecode analysis. Our experimental analysis shows that WANA can effectively detect vulnerabilities in both EOSIO and Ethereum smart contracts with high efficiency.

## CCS CONCEPTS

• **Security and privacy** → Software and application security

• **Software and its engineering** → Software verification and validation

## KEYWORDS

Symbolic execution, WebAssembly, Smart contract, Wasm, Vulnerability detection, EOSIO, Ethereum

## 1 Introduction

The blockchain technology has enabled decentralized value transfer networks among parties with limited trust [15]. With the support of smart contracts, the developers can build Decentralized Applications (DApps) on top of the blockchain platforms such that untrusted parties can cooperate with each other. Popular DApps include blockchain games, Decentralized Finance (DeFi), online gambling, decentralized exchanges, wallets, supply chain management, logistics tracking, etc.

[*] This research is supported in part by the National Key R&D Program of China under Grant 2019YFB2102400, and NSFC (project no. 61772056).
[#] WANA is being Open Sourced. https://github.com/gongbell/wana
[†] All correspondence should be addressed to Bo Jiang, gongbell@gmail.com.

The Ethereum [4] and EOSIO [29] are two of the most popular public blockchain platforms supporting smart contracts. However, the vulnerabilities within Ethereum and EOSIO smart contracts have led to financial loss to their end users. For EOSIO smart contracts, the Block Information Dependency vulnerability, the Fake EOS Transfer vulnerability and the Forged Transfer Notification vulnerability have led to the loss of around 380K EOS tokens [32][34] in total. The accumulated amount of loss by these vulnerabilities was around 1.9 million worth of USD at the time of attack. For the Ethereum smart contracts, the vulnerability in the DAO contract [23] led to the loss of $60 million. Also, the Freezing Ether and the Dangerous DelegateCall Vulnerability resulted in the loss of $60 million and the frozen of $150 million in terms of Ether [42][43], respectively.

Therefore, effective vulnerability detection tools are needed to safeguard the ecosystem of blockchain platform. However, the vulnerability detection tools for EOSIO smart contract are still limited. Furthermore, existing smart contract vulnerability detection tools can only support one blockchain platform. Due to the diversity of blockchain platforms, a cross-platform vulnerability detection tool is desirable for practical use.

WebAssembly (Wasm for short) [43] is a binary instruction format for a stack-based virtual machine. It is also adopted by the EOSIO blockchain platform for better efficiency and reliability. The Ethereum also plans to replace the EVM with Ethereum WebAssembly (EWasm) VM as the smart contract execution engine for Ethereum 2.0 [13][19]. As a result, the smart contracts on both the EOSIO and Ethereum platform will be compiled to WebAssembly for execution in the corresponding Wasm VM implementation. Therefore, a symbolic execution engine at the WebAssembly level has the potential to provide a common security analysis framework for EOSIO smart contract, Ethereum smart contract, and even for Web applications based on Wasm.

In this work, we proposed WANA, a symbolic execution engine for Wasm bytecode to support cross-platform vulnerability ANAlysis. WANA can support the vulnerability detection of both EOSIO and Ethereum smart contracts. In our experiment on about 3960 EOSIO smart contracts, WANA effectively detected more than 400 vulnerabilities. Furthermore, our case study showed that WANA can also effectively detect 3 typical vulnerabilities for Ethereum smart contracts.

The contribution of this work is three-fold. First, it presents the first cross-platform smart contract vulnerability detection framework based on the symbolic execution of Wasm bytecode. Second, it proposes a set of test oracles to detect the vulnerabilities within EOSIO and Ethereum smart contracts based on Wasm bytecode analysis. Third, the work has evaluated WANA on both EOSIO and Ethereum smart contracts for vulnerability detection, and WANA has effectively identified vulnerabilities in both EOSIO and Ethereum smart contracts with high efficiency.

The organization of the remaining sections is as follows. In Section 2, we present the basics of Wasm, EOSIO smart contracts and Ethereum smart contracts. Then in Section 3, we review typical vulnerabilities of EOSIO and Ethereum smart contracts. In Section 4, we present the design of our WANA framework in detail. After that, we report a comprehensive experimental study to evaluate the effectiveness of WANA in terms of vulnerability detection on EOSIO smart contracts in Section 5. Then we present a case study to evaluate the vulnerability detection ability of WANA on Ethereum smart contracts in Section 6. In Section 7, we present successful attacks to smart contracts detected by WANA. Finally, in Section 8 and Section 9, we present the related works and conclusions.

## 2 Background on Wasm and Smart Contracts

In this section, we will briefly review background information on Wasm and smart contracts.

### 2.1 The WebAssembly

WebAssembly (Wasm for short) [43][44] is a binary instruction format for a stack-based virtual machine. The WebAssembly virtual machines can be embedded into Web browsers or Blockchain platforms for execution. Wasm is currently supported by the EOSIO blockchain platform. And the replacement of Ethereum VM (EVM) with Ethereum Wasm VM is planned to take place in Ethereum 2.0 [13][19].

Wasm is designed as a portable target for compilation of high-level languages like C/C++/Rust, enabling deployment on the web for client and server applications. Wasm is designed to be fast, safe, and debugging friendly. As a result, Wasm is widely supported by all major web browsers [43]. There are two equivalent and convertible representations for WebAssembly: the binary format (".wasm" as suffix) and the text format (".wast" as suffix). The Wasm code is used for execution while the Wast code is for ease of reading and editing by human.

### 2.2 Background on EOSIO Smart Contracts

EOSIO [24] is a public blockchain platform that focuses on the scalability of transactions. The EOSIO platform uses the Delegated Proof of Stake (DPoS) consensus protocol which is much more scalable than the Proof of Work (PoW) consensus protocol. The EOS token also represents the stake hold by its owners. The EOSIO platform supports smart contracts to enable developers to build DApps.

On the EOSIO platform, there are two types of smart contracts. The first type is the system contracts, which are smart contracts deployed on the EOSIO platform by default to realize core blockchain features such as consensus, fee schedules, account creation, and token economics [40]. The *eosio.token* smart contract [31] is one such system smart contract which defines the data structures and actions enabling users to issue and transfer tokens on EOSIO based blockchains. The second type of smart contracts are user defined smart contracts deployed on the EOSIO platform [37] to realize specific business requirements.

Each EOSIO smart contract must realize the *apply* function as the entry function to handle actions. All incoming actions are routed to the *apply* function [27], which will in turn dispatch the actions to corresponding handler functions for processing. To develop EOSIO smart contract, the recommended language is C++ [29]. The smart contract in C++ will be compiled into Wasm bytecode for execution on the EOSIO Wasm VM. The *apply* function has several parameters as input: *receiver*, *code*, and *action*. The *receiver* is the account currently processing the action. The *code* is the account that the action was originally sent to. The *action* is the id of the action. When a smart contract receives an action, it may forward the action to other contracts with the *require_recipient()* function.

### 2.3 Background on Ethereum Smart Contracts

On the Ethereum blockchain platform, there are external accounts (i.e., owned by human) and smart contract accounts [34]. Smart contract is responsible to manage the balance and the persistent private storage of the contract account. A transaction is a message sent from one account to another account. Conceptually, Ethereum [32] can be viewed as a transaction-based state machine, where its state is updated on each transaction. When the target account of a transaction is a smart contract account, its code is executed and the payload is provided as input.

In existing releases of Ethereum, the smart contract code is executed on the Ethereum Virtual Machine (EVM). Developers can write smart contracts using Solidity, a high-level programming language [38], which are then compiled into EVM bytecode. In the upcoming releases of Ethereum 2.0, EWasm VM is planned to replace EVM to serve as the smart contract execution engine [13][19]. In particular, EWasm is a restricted subset of Wasm to be used for contracts in Ethereum. For example, it eliminates non-deterministic behaviors during smart contract execution.

## 3 Smart Contracts Vulnerabilities

In this section, we will review smart contract vulnerabilities in EOSIO and Ethereum blockchain platforms.

### 3.1 EOSIO Smart Contract Vulnerabilities

We will present 3 typical EOSIO smart contract vulnerabilities detected by WANA in this section.

#### 3.1.1 Fake EOS Transfer

In the EOSIO platform, a smart contract must send EOS tokens to another smart contract via the *eosio.token* system smart contract. The *eosio.token* system smart contract manages the EOS accounts of all smart contracts within its internal storage. During a transfer, the sender contract must call the *transfer* function of *eosio.token* system contract. Within the *transfer* function of *eosio.token*, the balances of the sender and receiver contract accounts will be adjusted accordingly. Meanwhile, *eosio.token* will call *require_recipient()* to notify both the *sender* and *receiver* contracts when performing the *transfer*.

Within the *apply* function of a safe smart contract, it must ensure the original receiver (i.e., the *code* parameter of *apply* function) of the *transfer* action is *eosio.token*. However, if the contract under attack is vulnerable in that it does not check whether the *code* is *eosio.token* within its *apply* function when the *action* is *transfer*, an attacker may perform an inline call to its *transfer* function directly to fake an EOS transfer. As a result, the vulnerable contract may wrongly consider that the attacker has transferred EOS to it. The EOSBet [25] and EOSCast [26] are two such contracts with Fake EOS Transfer vulnerability, and both have suffered serious losses.

| 1 | #define EOSIO_ABI_EX( TYPE, MEMBERS ) |
|---|---|
| 2 | extern "C" { |
| 3 | void apply(uint64_t receiver, uint64_t code, uint64_t action){ |
| 4 | auto self = receiver; |
| 5 | if (code == self || code == N(eosio.token) || action == N(onerror)){ |
| 6 | TYPE thiscontract (self); |
| 7 | switch( action ) { |
| 8 | EOSIO_API(TYPE, MEMBERS) |
| 9 | }}}} |

**Figure 1. The EOSBet Contract with Fake EOS Transfer Vulnerability**

As shown in Figure 1, within the *apply* function of the *EOSBet* contract, it only checks whether the code is the contract itself or *eosio.token* (line 5), but it does not check whether the code is *eosio.token* when the action is *transfer*. As a result, an attacker contract may directly call the *transfer* function of the vulnerable contract to make a bet without spending any EOS.

As shown in Figure 2, the best practice to fix the vulnerability is to add a check in the *apply* function to ensure the transfer is performed via the *eosio.token* system contract (lines 6 and 7). In other words, when the action is *transfer*, the code must be *eosio.token*.

| 1 | #define EOSIO_ABI_EX( TYPE, MEMBERS ) |
|---|---|
| 2 | extern "C" { |
| 3 | void apply(uint64_t receiver, uint64_t code, uint64_t action){ |
| 4 | auto self = receiver; |
| 5 | if( code == self || code == N(eosio.token) || action == N(onerror)){ |
| 6 | if( action == N(transfer) ){ |
| 7 | eosio_assert( code == N(eosio.token), "Must transfer EOS"); |
| 8 | } |
| 9 | … |

**Figure 2. The Fix of the Fake EOS Transfer Vulnerability**

#### 3.1.2 Forged Transfer Notification

During a Forged Transfer Notification attack [24], the attacker controls two accounts: *sender* and *notifier*. The *sender* initializes the attack by transferring EOS to *notifier* via *eosio.token*. When the transfer is successful, both *sender* and *notifier* will receive transfer notification. However, the smart contract with account *notifier* can deliberately forward the *transfer* action to a victim contract *C* with the *require_recipient()* function, which is essentially a carbon copy of the transfer action.

As shown in Figure 3, if the contract *C* fails to check whether the destination (i.e., *data.to*) of EOS transfer is itself within its *transfer* function, it may wrongly consider that it has actually received EOS from the *sender* rather than just a notification. As a result, it may wrongly credit the *sender* account, which in fact has sent nothing to *C*. The smart contract *C* is said to have the Forged Transfer Notification vulnerability [33].

| 1 | class C : public eosio::contract { |
|---|---|
| 2 | public: |
| 3 | void transfer(uint64_t sender, uint64_t receiver) { |
| 4 | auto data = unpack_action_data<st_transfer>(); |
| 5 | if (data.from == _self) //no check for data.to |
| 6 | return; |
| 7 | doSomething(); |
| 8 | } |
| 9 | } |

**Figure 3. The Smart Contract with Forged Transfer Notification Vulnerability**

As shown in Figure 4, a possible fix of the Forged Transfer Notification vulnerability is to check the destination of the transfer (line 5). If the destination of the EOS transfer is not the current contract, then the contract should directly ignore the transfer notification.

| 1 | class C : public eosio::contract { |
|---|---|
| 2 | public: |
| 3 | void transfer (uint64_t sender, uint64_t receiver) { |
| 4 | auto data = unpack_action_data<st_transfer>(); |
| 5 | if (data.from == _self || data.to != _self) |
| 6 | return; |
| 7 | doSomething(); |
| 8 | } |
| 9 | } |

**Figure 4. A Fix of the Forged Transfer Notification Vulnerability**

#### 3.1.3 Block Information Dependency

Reliable source of randomness is hard to obtain on blockchain platforms. Developers may be tempted to use the block information such as *tapos_block_prefix* and *tapos_block_num* to generate random numbers. The random numbers may be used to determine the transfer of EOS or the winner of a lottery. Unfortunately, the *tapos_block_prefix* and *tapos_block_num* are not reliable source of randomness because they can be directly calculated from *ref_block_num*, which is the id of the last irreversible block by default. A gambling contract may use a deferred action to determine the winner of a lottery. In such scenario, the reference block is the block just before the block making the bet. Therefore, when a smart contract uses *tapos_block_prefix* and *tapos_block_num* directly for random number generation, the number generated can be predicted.

EOSRoyale [32] is an EOSIO smart contract with Block Information Dependency vulnerability. As shown in Figure 5, it uses the product of the block number and block prefix as the seed for random number generation (lines 7 to 9). Therefore, the variables used for random number generation can be computed before making the bet. As a result, the attackers successfully calculated the random number, won the gambling game, and received the prizes in EOS.

| 1 | class EOSRoyale: public eosio::contract { |
|---|---|
| 2 | … |
| 3 | void rand() { |
| 5 | checksum256 result; |
| 7 | auto mixedBlock = tapos_block_prefix()* tapos_block_num(); |
| 8 | const char * mixedChar = (const char *)(&mixedBlock); |
| 9 | sha256((char*)mixedChar, sizeof(mixedChar), &result); |
| 10 | } |
| 11 | } |

Figure 5. The EOSRoyale Contract with Block Information Dependency Vulnerability

### 3.2 Ethereum Smart Contract Vulnerabilities

The Ethereum smart contract vulnerabilities [2][6][8][9][11][12][16][17][22] have been extensively studied in previous works. In this section, we will briefly review three common types of Ethereum smart contract vulnerabilities supported by WANA.

**Greedy.** A greedy smart contract can receive ether but it contains no functions to send ether to other accounts. Therefore, this kind of smart contract will freeze any ether sent to them. The second round of attack on Parity wallet vulnerability [42] was due to the fact that many smart contracts only relied on the parity wallet library to manage their ether through *delegateCall()*. When the parity library was changed to a contract through initialization and then killed by the attacker, all the ether within these wallet smart contracts relying on the parity library was frozen.

**Dangerous DelegateCall.** The delegatecall is similar to a message call except that the code is executed with the data of the calling contract [34]. This is the way to implement the "library" feature in Solidity for code reuse. However, when the argument of the *delegatecall* is *msg.data*, an attacker can manipulate the *msg.data* so that the attacker can make a victim contract to call a specific function. This vulnerability resulted in the outbreaks of the first round of parity wallet attack [41]. The Wallet contract used a *delegatecall* with *msg.data* as its parameter. As a result, an attacker could call any public function of *_walletLibrary*. Then the attacker called the *initWallet* function of the *_walletLibrary* and became the owner the wallet contract. Finally, the attacker sent the ether in the wallet to his own address to finish the attack, which led to $30 million loss to the parity wallet users.

**Block Information Dependency.** The Block Information Dependency vulnerability exists when an Ethereum smart contract uses the block timestamp or block number to determine a critical operation (e.g., sending ether or determining the winner of a lottery). Indeed, both *block timestamp* and *block number* are variables that can be manipulated by miners, so they cannot be used as reliable sources for critical operations. For example, the miner has the freedom to set the timestamp of a block within a small interval [36] in Ethereum. Therefore, if an Ethereum smart contract transfers ether based on timestamp, an attacker can manipulate the block timestamp to exploit the vulnerability.

## 4 The Design of the WANA Framework

In this section, we will present the design of the WANA symbolic execution framework for cross-platform smart contract vulnerability analysis.

### 4.1 The Workflow of WANA

The workflow of WANA is shown in Figure 6. The input of WANA is the Wasm bytecode. For EOSIO smart contract, the corresponding Wasm bytecode can be collected from the EOSIO public blockchain or compiled from EOSIO smart contract source code. For Ethereum smart contract written in Solidity, the corresponding bytecode can be generated from a solidity-to-Wasm compiler.

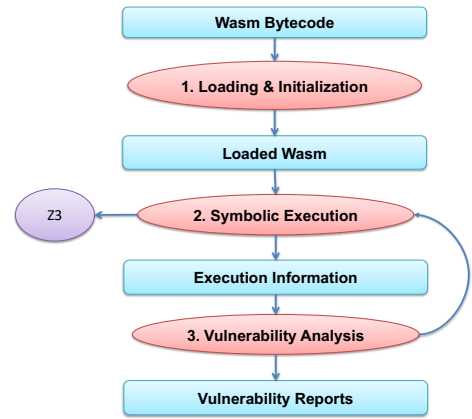

Figure 6 The Workflow of WANA

First, the Wasm bytecode is parsed, loaded and initialized within our Wasm symbolic execution engine. As a result, the stack and memory of the symbolic execution engine are ready for execution. Second, the symbolic execution engine starts to traverse the paths of the Wasm code with symbolic input. During symbolic execution, the symbolic execution engine will invoke the Z3 [14] constraint solver to check and prune unsatisfiable paths along the way. Meanwhile, the execution information useful for vulnerability analysis are collected during symbolic execution. Third, WANA will perform vulnerability analysis based on the execution information collected during the symbolic execution process. When a vulnerability is detected, the corresponding report will be generated. Note that the step 2 and step 3 are iterative process, where the vulnerability analysis and symbolic execution will interleave with each other. The whole workflow ends when all the code has been traversed.

## 4.2 The Symbolic Execution Engine

Wasm supports 4 data types, including two integer data types and two float-point data types under IEEE 754 standard. The WANA symbolic execution engine supports all of them for analysis.

WebAssembly programs are organized into *modules* [35], which are the units of deployment, loading, and compilation. For an EOSIO or Ethereum smart contract, the generated Wasm bytecode will correspond to one module. During the loading and initialization phase, the WANA symbolic execution engine will prepare the memories, tables, global variables and a stack as the execution environment for executing the Wasm instructions within a module. For a module, the *exports* component defines a set of exported functions and data structures that become accessible to the host environment once the module has been instantiated. In other words, the exported functions serve as the public interfaces for a Wasm module. Therefore, the WANA framework will start the symbolic execution by iterating each exported function.

Among the exported functions, the *apply* function is the entry function to dispatch actions for EOSIO smart contracts. For the Ethereum smart contracts, the *main* function is the entry function for interacting with the Ethereum smart contract Application Binary Interfaces (ABIs).

For each Wasm function invocation within the instructions in a module, WANA will first prepare a *frame* as its execution context, which includes arguments, local variables, return values, and references to its module. Then WANA will start symbolically executing the instructions within the code section of the function sequentially. The Wasm instructions mainly include the numeric instructions, memory instructions, control instructions, and function call instructions. WANA has realized 171 out of the 185 instructions in the WebAssembly specification version 1.0 [43].

For numeric instructions, WANA realizes the stack-based execution logic. The WANA symbolic execution engine will pop the operands, perform symbolic calculations, and push the result back to stack. All the numeric instructions on integer and floating-point value types are supported.

For control instructions, there are two kinds of branch instructions: unconditional branch (e.g., *br*) and conditional branch (e.g., *br_if*). The unconditional branch will directly jump to the label specified by the instruction argument to continue execution, which is useful to realize loops. However, when the depth of the loop grows, the symbolic execution engine may be very slow or trapped due to state explosion. Therefore, WANA set an upper bound to the depth of loops for each label during path exploration. For the instruction *br_if*, the execution will depend on the evaluation results of conditional expression. WANA will record the execution context for each branch and traverse each branch separately to realize path coverage. For each branch, the corresponding path constraint will be input to z3 [14] for constraint solving before traversing, which can improve the execution efficiency by pruning those infeasible paths.

There are three types of memory manipulation instructions: load, store, and increasing the size of the memory. Wasm adopts a linear memory model with a contiguous, mutable array of raw bytes [43]. A program can load and store values from/to a linear memory at any byte address. Since there are many bit manipulation instructions on integer and float data types in Wasm, we choose to represent the integer and float data types with the bit-vector data type in z3. WANA uses 32-bit (64bit) bit-vector data types to represent 32-bit (64-bit) integer and float-point symbolic data types. Since byte array cannot store bit-vector and its expression, WANA uses a linear list to store the references of both concrete and symbolic values to emulate memory.

In Wasm, there are two types of function calls: direct function calls and indirect function calls. For indirect function calls, Wasm VM has to first get an index from the top of the stack. Then the VM will use the index to get the real function address from a function table. There are 3 types of functions for calling directly: the other functions of the current Wasm module under analysis, the functions of other Wasm modules, the API functions of EOSIO. For the functions in the current Wasm module, WANA will step into the functions to continue symbolic execution. For the API functions or the functions in other modules, WANA still cannot support their symbolic executions. Therefore, WANA will randomly generate a concrete value based on the return type of the function, which may lead to sound but incomplete analysis. Modeling the functions in other contracts or the API functions may help improve WANA, which we left as a future work.

## 4.3 The Vulnerability Detection Algorithm

In this section, we will present the vulnerability detection algorithms for EOSIO and Ethereum smart contracts.

### 4.3.1 Vulnerability Detection for EOSIO Smart Contracts

**Fake EOS Transfer.** In general, to detect Fake EOS Transfer vulnerability, WANA checks whether there is a feasible path with *code* different from *eosio.token* that finally invokes the *transfer* function within the *apply* function.

| 1 | **bool** isFakeEOSTransfer() { |
|---|---|
| 2 | **if** *apply* function is reachable |
| 3 | **if** the *apply* function have 3 i64 parameters |
| 5 | find all feasible paths *P* in *apply* function where *code* is neither equal to *eosio.token* nor *receiver* |
| 6 | label the code invoking *transfer* function as *sink* |
| 7 | **if** there is any path in *P* reaching *sink* |
| 8 | **return true** |
| 9 | **return false** |

**Figure 7. Algorithms Detecting Fake EOS Transfer**

The algorithm to detect Fake EOS Transfer is shown in Figure 7. At line 2, WANA first checks whether the *apply* function is reachable. Then WANA further checks whether the *apply* function contains 3 64-bit integer parameters at line 3. The 3 64-bit integer parameters correspond to the *receiver*, *code*, and *action* parameters. In lines 5, WANA performs symbolic execution within the *apply* function to find all feasible paths *P* where code is neither equal to *eosio.token* nor *receiver*. In line 6, WANA labels the *wasm* code invoking the *transfer* function as *sink*. Finally, it considers the contracts vulnerable if there are any paths in *P* reaching *sink* (line 7). Otherwise, the contract is not reported as vulnerable (line 8 to 9).

WANA performs analysis on Wasm (i.e., binary format), but we use the WebAssembly text format (Wast) in the sequel of this section for ease of understanding. One possible Wast code pattern to compare the *code* parameter with *eosio.token* is shown in Figure 8. At line 3, the Wast code pushes $p1 (which represents the parameter *code*) to the stack. Then the Wast code pushes the 64-bit integer corresponding to *eosio.token* to the stack (line 5). Then, the *code* and *eosio.token* are compared at line 6. Line 7 represents the control flow branch when they are unequal and lines 9 to 11 represent the control flow segment when they are equal. The code paths in Figure 8 where *code* is not equal to *eosio.token* can be automatically identified during symbolic execution by adding the path constraint at the start: $p1 != encoding("eosio.token")$.

| 1 | ... |
|---|---|
| 2 | block $B2 |
| 3 | local.get $p1 //$p1 is code |
| 4 | //6138663591592764928 is i64 encoding of eosio.token |
| 5 | i64.const 6138663591592764928 //eosio.token |
| 6 | i64.ne |
| 7 | br_if $B2 //jump if not equal to eosio.token |
| 8 | // branch where code equals to eosio.token |
| 9 | i32.const 0 |
| 10 | local.set $l4 |
| 11 | br $B1 |
| 12 | end |

**Figure 8. WAST Code Comparing Code and eosio.token**

The Wast code identifying the invocation of the *transfer* function is shown in Figure 9. At lines 2 to 5, the Wast code compares the *action* parameter with *transfer*. Note that "$p2" represents the *action* parameter and the 64-bit integer at line 4 represents *transfer*. If the *action* equals to *transfer* (line 7), the Wast code will make preparations and call the *transfer* function (line 8 to 11). When detecting Fake EOS Transfer vulnerability, WANA uses the code pattern in Figure 9 to locate the code invoking *transfer* function (i.e., a sink).

| 1 | ... |
|---|---|
| 2 | local.get $p2 |
| 3 | //-3617168760277827584 is the i64 encoding of transfer |
| 4 | i64.const -3617168760277827584 |
| 5 | i64.ne |
| 6 | br_if $B10 //jump if not equal to transfer |
| 7 | //equal to transfer, so preparing the invocation of transfer |
| 8 | local.get $l3 |
| 9 | i32.const 0 |
| 10 | ... |
| 11 | call $f162 //invoking transfer |

**Figure 9. WAST Code Comparing Action and Invoking Transfer**

**Forged Transfer Notification.** When a smart contract receives a transfer notification, if it fails to check the destination (i.e., *data.to*) of EOS transfer is itself within its *transfer* function, a Forged Transfer Notification occurs.

The algorithm to detect Forged Transfer Notification is shown in Figure 10. At line 2, WANA first performs concrete execution on the *apply* function with the *action* and *code* parameters fixed to *transfer* and *eosio.token*, respectively. The goal of concrete execution is to locate the position of the *transfer* function. During the concrete execution, if the Wasm instruction *call_indirect* is executed, WANA will consider the *transfer* function reachable (line 3). This is because the *call_indirect* is used in Wasm of EOSIO smart contract to invoke ABI functions and the *action* parameter is fixed to *transfer* at the start. The *call_indirect* instruction will pop a 32-bit integer as index into a table storing function addresses to get the address of the *transfer* function in Wasm. Then WANA will record the position of the *transfer* function within Wasm for follow-up analysis (line 4).

| 1 | **bool** isForgedTransferNotification() { |
|---|---|
| 2 | concrete execution of *apply* with *action* fixed to transfer and *code* equals to *eosio.token*. |
| 3 | **if** *transfer* function is reachable: |
| 4 | label the position *pt* of *transfer* function in *Wasm* |
| 5 | label the variables or stack storing *_self* in *transfer* |
| 6 | label the *Wasm* comparing *_self* and *to* in *transfer* as *C* |
| 7 | symbolically executing the *transfer* function at *pt* |
| 8 | **if** there is any path reaching *C* |
| 9 | **return false** |
| 10 | else |
| 11 | **return true** |

**Figure 10. Algorithms Detecting Forged Transfer Notification**

From lines 5 to 11, the WANA analyzes the Wasm bytecode of the function *transfer*. At line 5, WANA first labels the set of variables storing *_self*, which is usually read from $p0 (the first parameter of *transfer*) and stored in local variables or the stack. Then it continues to label the *Wasm* bytecode comparing *_self* (in local variables or stack) with *to* as *C* (line 6). The "*to*" can be read from $p2 (the third parameter of *transfer*) or from the action data returned by *unpack_action_data()*. Finally, WANA performs symbolic execution within the *transfer* function to traverse all possible paths starting from *pt* (line 7). If there is any path reaching *C* during symbolic execution, the smart contract is considered not vulnerable (line 8 and 9). If there is no path reaching *C*, the smart contract is considered containing Forged Transfer Notification vulnerability (line 11).

**Block Information Dependency.** The algorithm to detect Block Information Dependency vulnerability is relatively straight forward and we will use the Wast representation for illustration. WANA first checks whether there is any invocation of the *$env.tapos_block_prefix* or *$env.tapos_block_num* functions to get the block information through static analysis. Then WANA further checks whether there are invocations to the *$env.send_inline* or *$env.send_deferred* functions to send EOS within the smart contract under analysis. Both the positions for collecting block information and for sending EOS are labelled during analysis. Finally, WANA performs symbolic execution to check whether there is at least one feasible path from the block information collection to the transfer of EOS. If there is one or more such feasible paths, the contract is reported to have Block Information Dependency vulnerability.

#### 4.3.2 Vulnerability Detection for Ethereum Smart Contracts

In this section, we will present the algorithms to detect the *Greedy*, *Dangerous DelegateCall*, and *Block Information Dependency* vulnerabilities in Ethereum smart contracts. For ease of understanding, we will also use the function names in the Wast representation to describe the vulnerability detection algorithm in this section. In Wasm, each function name corresponds to a 32-bit integer. The mapping between the function name and the 32-bit integer can be collected during smart contract loading within the symbolic execution engine.

**Greedy.** If an Ethereum smart contract can receive Ether but cannot send Ether, it is considered greedy. A greedy smart contract will freeze any Ether it has received.

To check whether an Ethereum smart contract can receive Ether, WANA checks whether there is at least one *payable* function reachable from the *main* function. WANA counts the total number of functions and the number of non-payable functions to calculate the number of payable functions. In particular, a non-payable function in Ethereum will use *$ethereum.getCallValue* to initialize, which is easy to identify.

To check that a smart contract cannot send Ether, WANA first checks whether there are any *$ethereum.call* functions within the bytecode of the smart contract. If there is not any *$ethereum.call* function within the bytecode of the smart contract, the contract cannot send any Ether. However, if there are some *$ethereum.call* functions, WANA will further check whether there is any feasible path from the *main* function to any of the *$ethereum.call* functions through symbolic execution. If there is no feasible path to any of them, the contract also cannot send any Ether.

**Dangerous DelegateCall.** To detect the Dangerous DelegateCall vulnerability, WANA checks whether the invocation of *DelegateCall()* function (i.e., *$ethereum.callDelegate* in the Wast) is reachable from the entry function and whether the input parameter specifying the function called by *DelegateCall()* is manipulatable by an attacking contract. To determine whether the function invoked is manipulatable, WANA checks whether the first argument (i.e., the function to invoke) of *$ethereum.callDelegate* is a constant value. If it is not a constant value, a vulnerability is detected.

**Block Information Dependency.** The detection of Block Information Dependency vulnerability for Ethereum smart contract is similar to that of the EOSIO smart contract. WANA also checks the existence of at least one feasible path from the block information collection to Ether transfer based on symbolic execution. The only difference lies in the implementation details. In Ethereum, the block information collection functions are *$ethereum.getBlockNumber*, *$ethereum.getBlockTimestamp* and *$ethereum.getBlockHash*, and the Ether transfer function is *$ethereum.call*.

## 5 Experiment on EOSIO Smart Contract Vulnerability Detection

In this section, we will present our experiment on EOSIO smart contract detection with WANA.

### 5.1 Research Questions

**RQ1**: Is WANA effective to detect the vulnerabilities within smart contracts?
**RQ2**: Is WANA efficient to detect the vulnerabilities within smart contracts?

### 5.2 Subject Programs

We collected 83 smart contracts with source code and 3881 smart contracts without source code for evaluation. The 83 smart contracts were downloaded from open source platforms. And the 3881 smart contracts were downloaded from distinct EOSIO accounts in EOSPark [27] and EOS Jungle [27] in September, 2019. The 3881 smart contracts accounted for a large percentage of contracts in distinct EOSIO accounts on the two websites.

### 5.3 Experiment Setup

We used a desktop PC as our experiment environment. The PC was running Ubuntu 18.04 and was equipped with Intel i7-6700 8-core CPU and 16GB of memory. The WANA tool was implemented by Python3 and we used the official Python release version 3.7. Finally, WANA used the z3 version 4.8.0 as the constraint solver. The default loop depth of WANA was set as 10 in the experiment.

### 5.4 Effectiveness of WANA

In this section, we present evaluation results to answer RQ1.

**Table 1. Vulnerability Detection Results on EOSIO Smart Contract with Source Code**

| Vulnerability | Total | WANA | | |
|---|---|---|---|---|
| | | Reported | FP | FN |
| Block Information Dependency | 83 | 3 | 0 | 0 |
| Forged Transfer Notification | 83 | 5 | 0 | 0 |
| Fake EOS Transfer | 83 | 2 | 0 | 0 |

**Results on smart contracts with source code.** The vulnerability detection results on EOSIO smart contracts with source code are shown in Table 1. We can see that among the 83 smart contracts, WANA reported 3 smart contracts with Block Information Dependency, 5 smart contracts with Forged Transfer Notification, and 2 smart contracts with Fake EOS Transfer. After manual checking the source code of the smart contracts, we confirmed that there were no false positives or false negatives for the 3 types of vulnerabilities by WANA in these 83 contracts.

Therefore, WANA is effective to detect vulnerabilities in EOSIO smart contracts with high accuracy in the experiment.

**Results on smart contracts without source code.** We further evaluated WANA on smart contracts without source code, the results are shown in Table 2. We can see that among the 3881 smart contracts, WANA has detected 1 Block Information Dependency vulnerability, 202 Forged Transfer Notification vulnerabilities, and 201 Fake EOS Transfer vulnerabilities. Among the 3881 smart contracts, the percentage of Block Information Dependency, Forged Transfer Notification, and Fake EOS Transfer are 0.026%, 5.2%, and 5.18%, respectively.

It is hard to manually verify those reported vulnerable smart contracts to calculate the precise number of false positives and false negatives. But the results give us an estimation of the number of vulnerable smart contracts deployed in the EOSIO platform in the wild. Given the relatively large percentages of Forged Transfer Notification and Fake EOS Transfer vulnerabilities, the developers are strongly recommended to perform security checks and harden their contract before releasing their EOSIO smart contracts.

**Table 2. Vulnerability Detection Effectiveness on EOSIO Smart Contracts without Source Code**

| Vulnerability | Total | Vulnerabilities Detected | Percentage |
|---|---|---|---|
| Block Information Dependency | 3881 | 1 | 0.026% |
| Forged Transfer Notification | 3881 | 202 | 5.20% |
| Fake EOS Transfer | 3881 | 201 | 5.18% |

## 5.5 Efficiency of WANA

In this section, we further analyze the vulnerability detection efficiency of WANA to answer RQ2. When the loop depth of WANA is set as 10 (default value), the smart contract analysis time is shown in Figure 11.

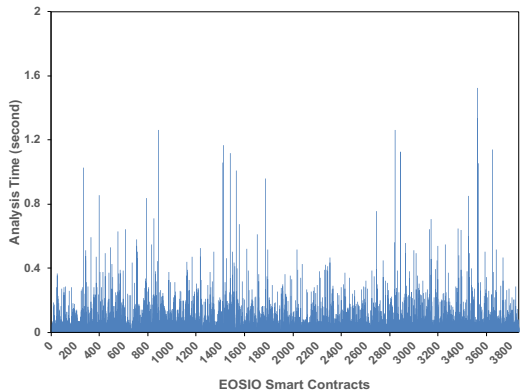

**Figure 11. Smart Contract Analysis Time of WANA**

Note the x-axis represents the 3881 smart contracts under analysis and the y-axis represents the analysis time for the corresponding smart contract. We can see that the smart contract analysis times range from 0.1 second to 1.6 second, which are quite practical. The average smart contract analysis time for EOSIO smart contract is 0.21 second. We further increased the loop count to 20 and 50, and the smart contract analysis times were all within 2 seconds. Therefore, WANA is efficient enough to perform vulnerability analysis on EOSIO smart contract for practical use.

## 6 Case Studies on Ethereum Smart Contracts Vulnerability Detection

In this section, we perform a case study on the detection of vulnerabilities for Ethereum smart contracts.

At the time of writing, the official support of Wasm in Ethereum is still under active development. As a result, there is no official toolchain to generate Wasm bytecode from Solidity or from EVM bytecode right now. But there are indeed ongoing projects working on Solidity-to-Wasm compilers. After manual checking, we found the SOLL compiler [39] is the only usable compiler, which can generate EWasm (Ethereum flavored WebAssembly) from Solidity and Yul.

However, the SOLL compiler still does not support library declaration, modifier declaration, multiple classes, contract inheritance, constructor with parameter, return multiple value, and Inline assembly at the time of our case study. Therefore, most of the Ethereum smart contracts cannot be successfully compiled into EWasm yet. We tried to compile around 15,000 Ethereum smart contracts with SOLL, but only less than 10 of them can be successfully compiled to Wasm. As a result, we cannot comprehensively evaluate WANA on a large number of Ethereum smart contracts. Therefore, we chose to manually modify a few typical vulnerable Ethereum smart contracts by removing unsupported features such that they can be supported by SOLL (without affecting their semantics and vulnerable behavior). Then we used these modified smart contracts to evaluate WANA.

**Table 3. Ethereum Smart Contract Used in Case Study**

| Vulnerability | Smart Contract Name |
|---|---|
| Greedy | Luck |
| | SaveData |
| | ShowNum |
| | TestRegistry |
| | whoDepositV3 |
| Dangerous DelegateCall | AdminInterface |
| | Delegatecall |
| | Router |
| | TransferReg |
| Block Information Dependency | AnyChicken |
| | EthMashMount |
| | Faucet |
| | TheOneToken |

## 6.1 Ethereum Smart Contracts Used in Case Study

We chose Ethereum smart contracts with 3 typical vulnerabilities supported by WANA to perform the case study:

Greedy, Dangerous DelegateCall, and Block Information Dependency. To make them usable with the version of SOLL at the time of writing (version 0.0.5), we manually modified the source code of each smart contract to remove unsupported syntax features. As shown in Table 3, we selected and modified 5 Greedy smart contracts, 4 Dangerous DelegateCall smart contracts, and 4 Block Information Dependency smart contracts for our case study.

### 6.2 Setup of the Case Study

We used the same desktop as in the experiment for EOSIO smart contract analysis to perform the case study. The desktop was installed with Ubuntu 18.04 as operating system. We used the SOLL version 0.0.5 to compile Ethereum smart contract to EWasm. And SOLL depended on llvm version 8.0.

When generating EWasm from Solidity, the SOLL compiler first analyzed the Solidity code to generate the LLVM IR (Intermediate Representation). Then the SOLL compiler continued to generate the EWasm bytecode from the LLVM IR. During our case study, we used WANA to analyze the EWasm bytecode to report any vulnerabilities.

### 6.3 Results and Analysis

The vulnerability detection results on Ethereum smart contracts are shown in Table 4. We can see that WANA has successfully identified all vulnerable smart contracts for each type of vulnerability. WANA produced no false negative cases in the case study. Due to poor compiler support at the current stage, we still cannot perform large scale experiment to comprehensively evaluate the accuracy of WANA on Ethereum smart contracts, which we left as a future work.

**Table 4. Vulnerability Detection Results on Ethereum Smart Contract**

| Vulnerability | WANA | |
|---|---|---|
| | Total | Reported |
| Greedy | 5 | 5 |
| Dangerous DelegateCall | 4 | 4 |
| Block Information Dependency | 4 | 4 |

We have also measured the efficiency of WANA on Ethereum smart contract. The smart contract analysis times ranged from 0.01 second to 0.53 second and the average smart contract analysis time was 0.08 second per smart contract. The results show that WANA is also efficient for analyzing Ethereum smart contracts.

## 7 Mounting Attacks on Vulnerable Smart Contracts

In this section, we manually mounted attacks on vulnerable EOSIO smart contracts detected by WANA.

### 7.1. Attacks on EOSIO Smart Contracts with Source Code

A previous version of *eosbetcasino* contract contained a Fake EOS Transfer vulnerability. And the contract was reported by WANA as vulnerable. We renamed the contract as *eosbetcasinopre* to differentiate it with other versions. Then we deployed and initialized the smart contract before mounting the attack.

We used an agent contract to perform the attack. The agent contract contained a function *fake*, which performed an inline call to the *transfer* function of *eosbetcasinopre*. The attack procedure mainly involved 3 steps:

In the first step, we invoked the function *fake* of the agent contract with *cleos*, which performed an inline call to the *transfer* function of the *eosbetcasinopre* to fake a bet.

*$ cleos push action agent fake [] -p agent@active*

In the second step, we queried the bet table of *eosbetcasinopre* to get the id of the current active bet. And a bet id value *14796464400348548551* was returned. In another word, we had successfully made a bet without paying any EOS.

*$ cleos get table eosbetcasinopre eosbetcasinopre activebets*

In the third step, we called the *refundbet* ABI function of *eosbetcasinopre* with the active bet *id* to get a refund before the bet was resolved. In reality, an attacker could choose to wait for the bet to resolve in the hope of winning a prize. In this attack, we asked for a refund to ensure the profit was locked in. Finally, we found the *fakeTransfer* account got a refund of 2.0 EOS but it paid nothing before.

*$ cleos push action eosbetcasinopre refundbet [14796464400348548551] -p eosbetcasinopre@random*

### 7.2 Attacks on EOSIO Smart Contracts without Source Code

The smart contract *pvpgamesrock* was reported by WANA to contain a Fake EOS Transfer vulnerability, so we proceeded to manually mount attacks on it.

First, we initiated an attack to call the *fake* function of the agent contract, which in turn made an inline call to the *transfer* function of the smart contract *pvpgamesrock*.

*$ cleos push action agent fake [] -p agent@active -f*

At first, we used a random value of EOS and memo strings to perform the inline call. However, the transaction was aborted and rolled back. Then we analyzed the strings referenced within the *transfer* function of *pvpgamesrock* through bytecode analysis and found the magic memo string "*malloc_from_freed was designed to only be called after _heap was completely allocated*" defined in the data area. Finally, we performed an inline call through the agent contract with the magic memo string to successfully made a bet without spending EOS. The parameters for the successful inline call was as follows:

```
["agent", "pvpgamesrock", "536.0402 EOS", "malloc_from_freed was designed to only be called after _heap was completely allocated"]
```

Then, we further called the *expireforce* function of *pvpgamesrock* to finish the attack by asking for a refund. We tried several strings for the parameter *server_str*, and we found any string starting with letter "f" worked. Therefore, we used the string "fire" to finish the attack. Finally, the smart *pvpgamesrock* transferred 536.0402 EOS as refund to our agent contract, who in fact spent no EOS to make the bet.

```
$ cleos push action pvpgamesrock expireforce '{"server_str": "fire"}' -p pvpgamesrock@active -f
```

## 8 Related Work

In this section, we present closely related work on smart contract vulnerability detection.

Atzei et al. [2] performed a comprehensive survey on vulnerabilities and attacks on Ethereum smart contracts. Parizi et al. [18] performed an experimental evaluation of current Ethereum smart contracts security testing tools. Chen et al. [5] proposed the TokenScope tool, which could analyze the transaction traces from Ethereum to automatically check whether the behaviors of the token contracts are consistent with the ERC–20 standards.

There are several works on the formal verification of smart contracts. Abdellatif and Brousmiche [1] proposed a formal modeling approach to verify the behavior of smart contract in its execution environment. They further performed security analysis on the smart contracts with a statistical model checking approach. Hirai [8] used Isabelle/HOL tool to verify the smart contract Deed, which was part of the Ethereum Name Service implementation.

Fuzzing is also an effective approach for vulnerability detection. Jiang et al. proposed ContractFuzzer [9], a black-box fuzzer for detecting vulnerabilities in Ethereum smart contracts. They also proposed the test oracles for detecting 7 typical Ethereum smart contract vulnerabilities. Nguyen et al. proposed a grey-box fuzzing tool called sFuzz [16] for smart contract vulnerability detection. The sFuzz tool adopts a lightweight multi-objective adaptive strategy to cover those hard-to-cover branches.

Symbolic execution [3][10] is a popular technique for vulnerability detection. There are several symbolic execution tools for security analysis of Ethereum smart contracts. Luu et al [12] proposed the Oyente symbolic verification tool for Ethereum smart contract. Oyente first builds the control-flow graph of smart contracts and then performs symbolic execution on the graph to check code patterns corresponding to smart contract vulnerabilities. Nikolic et al. [17] designed MAIAN, another symbolic execution tool to analyze the execution traces of Ethereum smart contracts. MAIAN characterizes the tracing properties of three typical smart contract vulnerabilities. And it can detect the greedy, the prodigal and the suicidal contracts in a scalable manner. Tsankov et al. proposed the Securify tool to detect Ethereum smart contract vulnerabilities. The Securify tool [20] first symbolically analyzes the contract's dependency graph to extract precise semantic information from the code. Then, it checks compliance and violation patterns that capture sufficient conditions for proving if a property holds or not.

Jhannes Krupp [11] proposed TETHER, a tool that automatically generates an exploit for an Ethereum smart contract given only its binary bytecode through symbolic execution. They also proposed an approach to handle hash values symbolically, which are used extensively in smart contracts.

ConsenSys proposed the Mythril [6] security analysis tool for EVM bytecode. Mythril used symbolic execution, SMT solving and taint analysis detect a variety of security vulnerabilities. Trailofbits proposed the Manticore [21] symbolic execution tool for dynamic binary analysis of smart contracts and Linux ELF binaries. Manticore combined symbolic execution, taint analysis, and instrumentation to analyze binaries.

Wang et al. [22] proposed the VULTRON tool based on the observation that almost all the existing transaction-related vulnerabilities were due to the mismatch between the actual transferred amount and the amount reflected on the contract's internal bookkeeping. The VULTRON tool provided a general test oracle that can be used to drive a range of smart contract analysis techniques.

EOSafe [7] is also a static analysis framework to detect vulnerabilities within EOSIO smart contracts based on symbolic execution. It can detect four types vulnerabilities in EOSIO smart contacts. Different from EOSafe, WANA aims at building a cross-platform smart contract vulnerable detection engine for both EOSIO and Ethereum platforms.

## 9 Conclusions and Future Work

The popularity of blockchain and smart contract technologies has enabled the cooperation among parties with limited trust. However, the vulnerabilities within smart contract have caused financial loss to its end users. Due to the diversity of the blockchain and smart contract platforms, existing vulnerability detection tools for smart contracts are generally platform specific. We observe that the smart contracts on both the EOSIO blockchain platform and the Ethereum blockchain platform will rely on WebAssembly VM for execution, which provides a common basis for building cross-platform smart contract vulnerability detection tools.

In this work, we proposed WANA, a symbolic execution engine for WebAssembly bytecode and a cross-platform smart contract vulnerability detection framework for both EOSIO and Ethereum smart contracts. The experiment and case studies on EOSIO and Ethereum smart contracts showed that WANA is effective and efficient to detect smart contract vulnerabilities.

The WebAssembly VM is also widely supported in Web browsers to support the development of Web application. Furthermore, the hardware-, language-, and platform-independent nature of WebAssembly also makes it popular for use in other environments. For future work, we plan to extend WANA to support the vulnerability detection of Web applications. We also plan to improve WANA to handle the symbolic analysis of external functions in a more accurate manner.